\documentclass{amsart}

\usepackage{amsmath}
\usepackage{amsthm}
\usepackage{amsfonts}
\usepackage{amssymb}
\usepackage{graphicx}
\usepackage{xcolor}
\usepackage{hyperref}
\usepackage{subcaption}
\usepackage{etoolbox}
\usepackage[foot]{amsaddr}
\usepackage{etoolbox}
\usepackage{multicol}
\usepackage{tcolorbox}
\usepackage{datetime}
    \newdateformat{monthyeardate}{%
  \monthname[\THEMONTH], \THEYEAR}
\usepackage{setspace}
\usepackage{longtable}
\usepackage[authoryear,compress]{natbib}
\usepackage{tikz}
\usetikzlibrary{calc}
\usetikzlibrary{decorations.pathreplacing}
\usetikzlibrary{patterns}
\usepackage{array}
\usepackage[utf8]{inputenc}

\hypersetup{
    colorlinks,
    linkcolor={red!50!black},
    citecolor={blue!50!black},
    urlcolor={blue!80!black}
}

\makeatletter
    \newtheoremstyle{indented}
        {12pt}% space before
        {12pt}% space after
        {\addtolength{\@totalleftmargin}{3.5em}
        \addtolength{\linewidth}{-3.5em}
        \parshape 1 3.5em \linewidth}% body font
        {}% indent
        {\bfseries}% header font
        {:}% punctuation
            {.5em}% after theorem header
        {}% header specification (empty for default)
\makeatother

\makeatletter
    \newtheoremstyle{indentedProp}
        {12pt}% space before
        {12pt}% space after
        {\addtolength{\@totalleftmargin}{3.5em}
        \addtolength{\linewidth}{-3.5em}
        \parshape 1 3.5em \linewidth}% body font
        {}% indent
        {\bfseries}% header font
        {:}% punctuation
            {0.5em}% after theorem header
        {}% header specification (empty for default)
\makeatother

\theoremstyle{indented}
    
\theoremstyle{indentedProp}
    
\theoremstyle{indented}
    
\theoremstyle{indented}
    
\theoremstyle{indented}
    
\theoremstyle{indented}
    
\theoremstyle{indented}

\makeatletter
    \patchcmd{\NAT@test}{\else \NAT@nm}{\else \NAT@nmfmt{\NAT@nm}}{}{}
    \DeclareRobustCommand\citepos
        {\begingroup
        \let\NAT@nmfmt\NAT@posfmt% ...except with a different name format
        \NAT@swafalse\let\NAT@ctype\z@\NAT@partrue
        \@ifstar{\NAT@fulltrue\NAT@citetp}{\NAT@fullfalse\NAT@citetp}}
    \let\NAT@orig@nmfmt\NAT@nmfmt
    \def\NAT@posfmt#1{\NAT@orig@nmfmt{#1's}}
\makeatother

\title[Metaethical Perspectives on `Benchmarking' AI Ethics]{Metaethical Perspectives \\ on `Benchmarking' AI Ethics}

\author{Travis LaCroix$^1$ \\ Alexandra Sasha Luccioni$^2$}
\address{$^1$Department of Philosophy \\ Dalhousie University}
\email{$^1$tlacroix@dal.ca}

\address{$^2$Hugging Face}
\email{$^2$sasha.luccioni@huggingface.co}

\date{Unpublished draft of \monthyeardate\today. {\it Please cite published version, if available}}

\begin{document}

\maketitle

\begin{abstract}
\singlespacing
    Benchmarks are seen as the cornerstone for measuring technical progress in Artificial Intelligence (AI) research and have been developed for a variety of tasks ranging from question answering to facial recognition. An increasingly prominent research area in AI is ethics, which currently has no set of benchmarks nor commonly accepted way for measuring the `ethicality' of an AI system. In this paper, drawing upon research in moral philosophy and metaethics, we argue that it is impossible to develop such a benchmark. As such, alternative mechanisms are necessary for evaluating whether an AI system is `ethical'. This is especially pressing in light of the prevalence of applied, industrial AI research. We argue that it makes more sense to talk about `values' (and `value alignment') rather than `ethics' when considering the possible actions of present and future AI systems. We further highlight that, because values are unambiguously relative, focusing on values forces us to consider explicitly {\it what} the values are and {\it whose} values they are. Shifting the emphasis from ethics to values therefore gives rise to several new ways of understanding how researchers might advance research programmes for robustly safe or beneficial AI. We conclude by highlighting a number of possible ways forward for the field as a whole, and we advocate for different approaches towards more value-aligned AI research.
    
    \phantom{a}

\end{abstract}

\setcounter{page}{1}

\section{Introduction}

    Benchmarks are a key tool for measuring technical progress in artificial intelligence (AI) research. A variety of benchmark datasets have been developed to measure a model's performance on particular tasks, such as question answering~\citep{Rajpurkar-et-al-2016}, facial recognition~\citep{Huang-et-al-2008}, machine translation~\citep{Bojar-et-al-2014}, etc. At the same time, the subject of AI ethics---including questions surrounding safety, fairness, accountability, transparency, etc.---has become increasingly prominent as a research direction in the field in recent years. However, there is presently no community-accepted standard for measuring the `ethicality' of an AI system---i.e., whether the decisions rendered by an AI system are morally `correct'. That is to say, there is no {\it benchmark} for measuring whether an AI system `is' ethical or for comparing the performance (in morally-loaded scenarios) between two distinct models or use cases. %

    In this paper, drawing upon research in moral philosophy---including normative ethics and meta-ethics---we argue that it is, in fact, impossible to develop such a benchmark. Part of the problem arises because the word `ethics' carries significant philosophical and conceptual baggage. Furthermore, members of the AI community are not always sensitive to the subtleties and problems that drive research in moral philosophy. For example, some researchers have suggested that moral {\it dilemmas}---a type of philosophical thought experiment---may be useful as a verification mechanism for whether a model chooses the ethically-`correct' option in a range of circumstances. But, these dilemmas, in the context of benchmarking ethics, often fail to maintain sensitivity to, e.g., the purpose of philosophical thought experiments like moral dilemmas~\citep{LaCroix-2022-Moral-Dilemmas}. Further problems arise because of the implicit assumptions that AI researchers make about the very nature of ethics---particularly, meta-ethical assumptions about the {\it objectivity} of ethics. These insights help clarify why attempts to benchmark ethics for AI systems presently fail and why they will continue to do so.

    Thus, we argue that alternative mechanisms are necessary for evaluating whether an AI system `is' ethical. These considerations are especially pressing in light of the prevalence of applied industrial AI research. We argue that it makes more sense to talk about `values' (and `value alignment') rather than `ethics' when considering the possible actions of present and future AI systems. We further highlight that because values are unambiguously relative, focusing on values rather than ethics forces us to consider explicitly {\it what} and {\it whose} values they are. This practice has additional downstream benefits for conceptual clarity and transparency in AI research. Therefore, shifting the emphasis from ethics to values gives rise to several new ways of understanding how researchers might move forward with a programme for robustly safe or beneficial AI. %

    We begin with a discussion of benchmarking in general, highlighting some of the issues recently identified for existing machine learning (ML) datasets and benchmarks (Section~\ref{sec:Measuring-Progress}). We then consider benchmarks in the context of ethics for AI systems (Section~\ref{sec:Moral-Benchmarks}) and why they fail. In particular, we discuss a supposed benchmark for ethical AI that has arisen in the context of autonomous vehicles as a particular case study: the `Moral Machine Experiment' (MME)~\citep{Moral-Machine}. We follow with a discussion regarding what values are transmitted via AI research and whose values they are (Section~\ref{sec:ethicalvalues}). We conclude by highlighting several possible ways forward for the field as a whole, and we advocate for different approaches towards more ethically-aligned AI research (Section~\ref{sec:Ways-Forward}). % {\color {red} {\bf Note}: Provide summary of `ways forward', once we know what we are saying.}

\section{Measuring Progress in Artificial Intelligence}
    \label{sec:Measuring-Progress}

     Generally speaking, a benchmark can be described as a dataset in combination with a metric---defined by some set of community standards--- used for measuring the performance of a particular model on a specific task~\citep{Raji-et-al-2021}. Benchmarks are meant to provide a fixed and representative sample for comparing models' performance and tracking `progress' on a particular task. In this section, we describe some examples of benchmarking results for typical ML tasks and then highlight the myriad ways that have been noted in the literature in which these standard benchmarks give rise to certain issues (\ref{sec:Benchmarking-Problems}). We then discuss how human performance on certain tasks is increasingly used to benchmark model performance and why this approach is illogical given the differences between humans and algorithms (\ref{sec:Benchmarking-Humans}).

\subsection{Issues with Existing Benchmarks}
    \label{sec:Benchmarking-Problems}

    Since its inception, designing tasks and measuring model performance have been central to the field of AI. These continue to be an important part of how members of the AI community measure `progress'. However, despite the ubiquity of benchmarking, major issues have been identified in existing ML datasets and benchmarks.\footnote{In the context of this paper, we use `AI' to refer to general approaches in the field pursuing machine intelligence, and we use `ML' to refer specifically to non-linear statistical approaches within that field.} These issues can arise from, e.g., subjective or erroneous labels~\citep{Gebru-et-al-2018} or a lack of representation, leading to systematic failures across datasets and evaluative approaches~\citep{Raji-et-al-2021, Liao-et-al-2021}. For example, datasets like ImageNet~\citep{Deng-et-al-2009} depend on linguistic hierarchies created in the 1980s and include outdated terms such as `harlot' and `chimneysweep'~\citep{Crawford-Paglen-2019}. At the same time, some of the most commonly-used datasets (including ImageNet) have been shown to contain an average of $3.3$\% labelling errors, with some datasets having error rates up to $10$\%~\citep{Northcutt-et-al-2021}.%
            %%%
            \footnote{If $3.3$\% sounds like a reasonable error rate, consider that these datasets are often huge. ImageNet contains more than $14$ million images, meaning that nearly $1$ million of these images---commonly used for training---might be erroneously labelled.}

    At best, these issues can affect model performance since they represent noisier data, making it harder for models to learn meaningful representations~\citep{Reed-et-al-2015} and for researchers to evaluate model performance properly~\citep{Northcutt-et-al-2021}. Further, this can preserve problematic stereotypes or biases, which are difficult to identify in models deployed in the real world~\citep{Koch-et-al-2021, Yang-et-al-2020}. At worst, they may reinforce, perpetuate, and even {\it generate} harms by creating negative feedback loops that further entrench societal structural inequalities \citep{O-Neil-2016, Falbo-LaCroix-2021}.

    Above and beyond specific datasets, entire AI tasks---such as recognising faces and emotions---have been repeatedly flagged as problematic (often for similar reasons as described above)~\citep{Buolamwini-Gebru-2018, Stark-2018, Stark-Hutson-2021}. Nonetheless, these tasks continue to be used for benchmarking models and developing entire systems. One such task involves predicting the mortality of different passengers aboard the HMS Titanic~\citep{Kaggle-2012}. This task has been used for hundreds of tutorials, blog posts, and ultimately published studies.%
        \footnote{See, for example, \citep{Kakde-Agrawal-2018, Barhoom-et-al-2019, Singh-et-al-2020, Tabbakh-et-al-2021, Shekhar-et-al-2021}.} 
    However, whether or not a particular passenger survived is mostly predicted by their {\it gender} and the fare they purchased---i.e., their {\it class} or {\it social status}~\citep{Broussard-2018}. So, the task of predicting the fate of passengers on the Titanic is perhaps morally dubious---especially when it is done without considering the social inequalities that gave rise to differential mortality rates in the first place. 
    
    Consider another example, from the field of computer vision. Oft-used tasks have included applying makeup to images of female faces~\citep{Jiang-et-al-2020, Li-et-al-2018, Chang-et-al-2018}, changing women's clothes from pants to mini-skirts~\citep{Mo-et-al-2018, Yang-et-al-2014}, and censoring nude women's bodies by, e.g., covering breasts with a bikini top~\citep{Simoes-et-al-2019, More-2018}. Such tasks are ethically problematic because they perpetuate gendered biases and stereotypes, thus reinforcing harmful systems of sexism and misogyny~\citep{Manne-2018}. Even so, these tasks are routinely used as acceptable benchmarks for computer vision models and their results are accepted at leading AI conferences, such as CVPR and ICCV.

    Although some publication venues---academic conferences and journals---are starting to forward ethical guidelines for both authors and reviewers~\citep{The-Other-Bengio-et-al-2021}, there is still a general lack of consensus about what constitutes acceptable tasks and applications of ML. This variance exacerbates the fact that it is not obvious that such guidelines will be effective in the first place~\citep{LaCroix-Mohseni-2020}. Furthermore, creating larger and larger datasets is relatively cheap, but the process of filtering those datasets or `detoxifying' the models trained on them is expensive \citep{Birhane-et-al-2021-Multimodal, Welbl-et-al-2021, Xu-et-al-2021}. In addition, even when these changes in the direction of `more ethical' or for a `common good' are well-intentioned, the lack of conceptual clarity surrounding the targets of such change---i.e., considering what it means to `be ethical' in the first place---will only compound the issue~\citep{Taylor-2016, Green-2019, Moore-2019, Cowls-2021}.

    In natural language processing (NLP), issues with benchmarks can be more subtle to identify. Still, these may range from unscientific task framing (such as predicting IQ scores based on written text~\citep{Johannssen-et-al-2020}) to embedded gender and cultural stereotypes in common NLP benchmarks~\citep{Blodgett-et-al-2021}. For example, in a recent survey of gender biases in NLP models,~\citet{Stanczak-Augenstein-2021} highlight four key limitations for NLP research:%
            %%%
            \footnote{In this context, biases can be understood as behaviours that involve {\it systematic discrimination} against specific individuals or groups (typically in favour of other individuals or groups)~\citep{Friedman-Nissenbaum-1996}.} 
            %%%
    (1) gender is often interpreted in a binary fashion, leading to, e.g., misgendering or erasure of non-binary gender identities~\citep{Behm-Morawitz-Mastro-2008, Fast-et-al-2016}; (2) NLP research is primarily monolingual, often focusing solely on the English language~\citep{W3TechS, Koroteev-2021, Wang-et-al-2019, Conneau-Kiela-2018}; (3) biases are typically tested {\it post hoc}---i.e., after the model has been deployed~\citep{Mitchell-et-al-2019}; and (4) when research explicitly tests for bias (which is infrequent), the evaluation metrics are often incoherent~\citep{Stanczak-Augenstein-2021}. Thus, even when benchmarks exist for a particular task, researchers lack good baselines for testing ethics considerations in their models---of which bias is one salient example. However, most newly-developed algorithms in this field do not test their models for biases in the first place, and ethical considerations are often ignored.

     %Another line of research on benchmarking has focused on the way in which we measure `progress', which is taken to be important given the heterogeneity of metrics currently used for benchmarking ML systems. When proposing a new model or architecture, comparing the model's performance on a relevant benchmark (or, in some cases, set of benchmarks) is seen as essential for providing evidence of improved performance.  In fact, one recent analysis of almost 4000 ML papers has found that more than two thirds only track a single performance metric~\citep{Blagec-et-al-2021}. Increasingly, model performance is being compared to human performance on particular tasks.  But, such metrics often fail to correlate with human judgement~\citep{Blagec-et-al-2021}.     Thus, benchmarking tasks is often done unreflectively, and this gives rise to myriad technical, social, and conceptual problems. 

\subsection{Benchmarking Humans and Machines}
    \label{sec:Benchmarking-Humans}

    As mentioned, AI models' performance is increasingly compared to that of humans, with some models reporting `superhuman performance' on, e.g., game-playing~\citep{Tesauro-1995, Schaeffer-et-al-1996, Campbell-et-al-2002, Mnih-et-al-2013, Silver-et-al-2016, Brown-Sandholm-2017, Moravcik-et-al-2017}, image recognition~\citep{He-et-al-2015}, NLP tasks~\citep{He-et-al-2021}, etc. However, such comparisons are often misguided (at best) and incoherent (at worst). Recent research has shown that many `superhuman' language models fail on simple challenge examples requiring compositionality~\citep{Nie-et-al-2019}, logical reasoning~\citep{Glockner-et-al-2018}, or even simple negation~\citep{Hossain-et-al-2020}. At the same time, human performance on certain tasks---e.g., diagnoses from X-rays---are often measured by the accuracy of binary outputs---a particular diagnosis is either positive or negative. In contrast, diagnostic AI models are continuous, including certainty or confidence \citep{Gichoya-et-al-2018}---this makes it difficult to compare the two, since the decision threshold can change depending on model parameters. Finally, comparing human and machine performance using the same metrics is precarious because metrics such as accuracy, widely used in AI, often fail to correlate with human judgement~\citep{Blagec-et-al-2021}.  Thus, there is a sense in which human performance on tasks is incomparable to computer performance, making any claim of comparison incoherent---not to mention that such comparisons imply `a narcissistic human tendency to view ourselves as the gold standard'~\citep{Lee-2021}.

    %This schism between the ever-improving performance of models on canonical datasets and the brittleness of these same models when faced with real-world phenomena can be explained by Goodhart's law: benchmarks have become both a {\it target} and a {\it measure}. This incentivises model behaviour that optimises performance as opposed to criteria like robustness and generalisability. {\color {red} [Sentence or two making explicit why this is bad.]}

    But given this divergence, it is important to systematically measure progress in AI, either alone or in comparison with `human-level performance'. However, for this to be possible, performance metrics should provide similar conditions for humans and algorithms. An emerging research topic seeks to bridge this gap by establishing more `equitable' settings for such comparisons---e.g., by imposing constraints such as reduced exposure time for algorithms~\citep{Funke-et-al-2021} or a restricted set of label options for humans~\citep{Dujmovic-et-al-2020}. For instance, recent work shows that running images through human-like processing filters before feeding them through an algorithm helps even the playing field for both humans and machines~\citep{Elsayed-et-al-2018}. These insights have led to proposals that AI models' performance on standard benchmarking tasks is not representative of any underlying capacity or lack thereof, given the nature and context of the tasks~\citep{Firestone-2020}.

    Another way of making the human-machine comparison more coherent is developing ways for models to signal that they do not know how to solve, for instance, a classification task. This ability would require developing new ways for quantifying and integrating uncertainty into the decision-making process since current approaches do not provide an `I don't know' option in the categories available during classification. In real-world deployments of AI systems, this behaviour is often addressed with heuristics (e.g., a cutoff based on a logit below a certain value). However, this does not solve the underlying issue that existing models do not know when they do not know;\footnote{Some philosophers have suggested that understanding the limits of knowledge is a prerequisite for {\it wisdom} (as opposed to mere intelligence)~\citep{Fucking-Socrates-ever-heard-of-him?}.} this makes it difficult to compare human and machine classification processes.
    
    %This is both inherently flawed and ultimately unrepresentative of the way in which humans approach similar tasks.

    Existing proposals have forwarded new evaluation benchmarks that aim at measuring models' robustness and capacity to generalise to new tasks, both from a natural language~\citep{Yogatama-et-al-2019, Bowman-Dahl-2021, Clark-et-al-2018} and a computer vision perspective~\citep{Hendrycks-Dietterich-2018, Mu-Gilmer-2019}, finding that many models that succeed at existing benchmarks fail at these. Recent work has also proposed alternative approaches such as iterative benchmark development~\citep{Ettinger-et-al-2017} and dynamic benchmarking~\citep{Kiela-et-al-2021}, which endeavour to bring entire fields towards a more nuanced, complex, and informed way of comparing models and measuring progress~\citep{Denton-et-al-2020, Schlangen-2020}. However, even if the issues with existing benchmarks (and their underlying datasets) on well-defined tasks are resolved, these problems severely limit any possibility of benchmarking ethics for AI systems insofar as ethics tasks are rarely, if ever, well-defined. This difficulty is a consequence of the very nature of ethics, as we discuss in the next section.

\section{Moral Benchmarks for AI Systems}
    \label{sec:Moral-Benchmarks}

    As AI systems become increasingly autonomous and more deeply integrated with society, it is obvious that some of the decisions made by these systems will begin to have moral weight. For example, consider a narrow chess-playing algorithm that can only make decisions confined to the action space provided by a chessboard. If the model `decides' to open with the Queen's Gambit, this is not a {\it moral} decision under any definition of `morality'. In contrast, the decisions made by an autonomous weapon system~\citep{Arkin-2008a, Arkin-2008b, Krishnan-2009, Tonkens-2012, Hellstrom-2013, Asaro-2020}, a healthcare robot~\citep{Anderson-et-al-2006, Anderson-Anderson-2008, Sharkey-Sharkey-2012, Conti-et-al-2017}, or an autonomous vehicle~\citep{Bhargava-Kim-2017, Sommaggio-Marchiori-2018, Evans-et-al-2020} may have moral weight. In these cases, the action space may include decision points that we might call `moral' or `immoral'---for example, choosing to prioritise one patient over another. %for example, choosing to prioritise one patient over the other, or choosing to initiate a drone strike on a building given that it contains a terrorist based on surveillance footage (and a certain confidence threshold)

    Part of the distinction between a chess-playing algorithm, whose decisions are confined to a particular action space, and an algorithm that acts in the real world is that the decisions made by the latter systems have the potential to impact others. So, in theory, deploying AI systems in the real world logically implies that they will sometimes need to make decisions with moral weight. However, as the action space increases, the set of possible failure modes increases exponentially.  Further, the economic promise of AI implies that these systems {\it are} increasingly being deployed in society rather than being rigorously tested in the confines of a research lab, thus increasing the risk of harm~\citep{LaCroix-Bengio-2019, Luccioni-Bengio-2020}. Of course, it is not necessary to posit some future science-fiction version of an AI robot acting autonomously in the world to see that the decisions of AI systems may create harm. As a case in point, even narrow AI systems today perpetuate harmful biases, affecting real-world outcomes~\citep{Angwin-et-al-2016, Christian-2020, Tomasev-et-al-2021}. And, as mentioned, these decisions may give rise to negative feedback loops, which further entrench those biases (and the harms caused by them) in society~\citep{O-Neil-2016, Falbo-LaCroix-2021}.

    It should come as no surprise, then, that research on ethical behaviour or decision-making in AI systems would attempt to construct a coherent measure for determining whether a system is `acting ethically'---i.e., whether the decision the model renders is morally `correct'. Given the historical importance of benchmarks for developing and evaluating AI systems, it makes sense that researchers would try to utilise this tool for evaluating the {\it moral} performance of an AI system. However, we argue in this section that benchmarking ethics in this way is impossible. First, we highlight how AI researchers have used moral dilemmas from philosophy as benchmarks for moral performance (\ref{sec:MoralDilemmas}) and some recent work criticising this approach (\ref{sec:MoralMachines}). We then introduce philosophical research in metaethics to show why it is impossible to benchmark ethical behaviour (\ref{sec:Meta-Ethics}). Finally, we turn our discussion toward real-world distributions to highlight that even if our claims about the nature of ethics turn out to be false, it will still be impossible to benchmark ethical behaviour in an AI system (\ref{sec:LongTails}).

    Thus, the arguments of this section are primarily negative. However, in the subsequent section, we provide a positive argument in favour of shifting the discourse of AI ethics toward talk of {\it values}. We discuss how such a shift would avoid many of the problems to which attempts to benchmark ethics give rise.

\subsection{Moral Dilemmas and Normative Theories}
    \label{sec:MoralDilemmas}

    The most common metric for evaluating whether or not a system `is' ethical is how the algorithm performs on particular moral {\it dilemmas}~\citep{Nallur-2020}. Before we discuss benchmarking ethics {\it using} moral dilemmas, we introduce what a moral dilemma is in the first place. To take a concrete example, trolley-style problems are sometimes used to consider certain morally-loaded decisions that autonomous vehicles (AVs) might have to make as these systems become increasingly ubiquitous in society. The trolley problem was originally introduced by Philippa Foot~\citep{Foot-1967}---and later extended by Judith Jarvis Thomson~\citep{Thomson-1976, Thomson-1985}---to consider why it might be permissible to perform some intentional action, $A$, in situation, $S$, despite its foreseeable (and undesirable) consequences.\footnote{This principle dates to at least~\citet{Aquinas-1485}; Foot calls it the {\it Doctrine of Double Effect}~\citep{FitzPatrick-2012}. See also discussion in~\citep{Kamm-1989, Unger-1996}.} Consider the following scenario.
    %%%%%
        \begin{quote}
        \singlespacing
            {\bf Bystander at the Switch} \\
            Suppose there is a trolley heading toward five individuals tied up on the tracks and unable to move. You are near a switch, which would divert the trolley to a separate track, where there is only one individual on the track (also unable to move). You have two (and only two) options:
            \begin{quote}
                \begin{enumerate}
                    \item Do nothing, in which case the trolley is guaranteed to kill the five people on the main track.
                    \item Pull the switch, diverting the trolley onto the side track where it is guaranteed to kill one person.
                \end{enumerate}
            \end{quote}
        \end{quote}
    %%%%%
    This standard formulation can be contrasted with the following alternative trolley problem:
    %%%%%
        \begin{quote}
        \singlespacing
            {\bf Bystander on the Footbridge} \\
            Suppose you are on a footbridge above a set of trolley tracks. Below, an out-of-control trolley is approaching five people on the track. The only way to stop the trolley is by dropping something of sufficiently heavy weight onto the tracks to block its path. As it happens, there is a person nearby of sufficiently heavy weight. You have two (and only two) options:
            \begin{quote}
                \begin{enumerate}
                    \item Do nothing, in which case the trolley will kill the five people on the track.
                    \item Push the person off the bridge, thus killing them (but thereby saving the five others).
                \end{enumerate}
            \end{quote}
        \end{quote}
    %%%%%
    Each of these is a particular type of {\it philosophical thought experiment}, called a {\it moral dilemma}~\citep{McConnell-2018}. Note that different {\it normative theories} from moral philosophy might offer divergent {\it prescriptions} (or {\it proscriptions}) when these two cases---{\bf Switch} and {\bf Footbridge}---are considered together. In this context, `normativity' concerns an evaluation or judgement---e.g., that one {\it ought} to do something. (We will use the phrase `normative theory' throughout this paper to refer to theories from moral philosophy, without necessarily committing to any claims about `morality' or `ethics'.) A `prescription' can be understood as the provision of a rule to follow or an action to take---i.e., a prescription that one {\it ought} to $\phi$ or that one {\it must} $\phi$. In contrast, a `proscription' is the provision of something forbidden---i.e., a proscription that one {\it ought not} to $\phi$, or that one {\it must not} $\phi$.

    Consider a concrete example of how distinct normative theories may offer divergent prescriptions in the same scenario. Certain forms of utilitarianism~\citep{Mill-1863, Bentham-1789}---a consequentialist normative theory that prescribes utility-maximisation as a reason for action---would recommend acting in {\it both} {\bf Switch} {\it and} {\bf Footbridge} because five deaths are obviously worse than one death. On the other hand, a Kantian brand of deontology~\citep{Kant-1785, Korsgaard-1996, Korsgaard-2009}---a non-consequentialist normative theory which emphasises the importance of duties---would at least say that it is impermissible to act in {\bf Footbridge} since this requires treating a human agent as a means to an end, rather than an end in itself, thus violating the {\it Categorical Imperative}~\citep{Kant-1785}.%
            %%%
            \footnote{The categorical imperative states it is never permissible to use a human agent as a means to an end. It is less obvious whether this imperative would also proscribe acting in {\bf Switch}. However, \citet{Thomson-1985} argues that there is a sense in which {\bf Switch} still uses a human agent as a means to an end and thus would be impermissible by Kantian deontology.} 
            %%%
    So, two different normative theories may prescribe (or proscribe) different actions in the same context because they take competing considerations to be important for moral decisions---in this example, consequences on the one hand and duties on the other.

    In many cases, different normative theories will prescribe the same action (although, possibly for different reasons). However, as we have seen, there may be some tension between the prescriptions of these theories, and moral dilemmas can serve to make these differences salient. Further, moral dilemmas underscore tensions between individual intuitions regarding the rightness or wrongness of an action in a given scenario. In empirical studies, most individuals say they would only act in the case of {\bf Switch}, not in {\bf Footbridge}~\citep{Navarrete-et-al-2012, Bourget-Chalmers-2014}. Thus, both the prescriptions of normative theories {\it and} common intuitions about the permissibility of an act may vary.%
            %%%
            \footnote{Of course, how people respond to abstract philosophical dilemmas on questionnaires may be quite different from how they act in the real world~\citep{Navarrete-et-al-2012, Bostyn-et-al-2018}.} 
            %%%
    The point is that a moral dilemma is a tool for philosophical analysis used to bring these tensions to the fore.

    Part of the purpose of a moral dilemma (as a type of philosophical thought experiment) is to focus attention on the morally-{\it salient} features of the dilemma~\citep{Dennett-1984, Dennett-1992, Dennett-2013, Brown-Fehige-2019} without getting bogged down by the pre-theoretic baggage that individuals may carry. In the case of the trolley problem, Foot's original target of analysis is {\it abortion} (not trolleys)~\citep{Foot-1967}. However, the thought experiment is useful precisely because of the supposed tension (at least in western analytic philosophy) between {\it emotion} and {\it rationality}~\citep{James-1890, Jagger-1989, Spelman-1989, Fricker-1991}: people are less likely to carry pre-theoretic baggage about trolleys than they are about abortions. Therefore, the thought experiment gets to the core of a moral issue in applied ethics while abstracting away from the actual (morally-loaded) target~\citep{LaCroix-2022-Moral-Dilemmas}. %
    Despite the conceptual purpose of dilemmas in moral philosophy---i.e., as thought experiments or `intuition pumps'~\citep{Dennett-1984, Dennett-1992, Dennett-2013}---AI researchers have begun to use these dilemmas as validation proxies for whether an model is ethical. In the remainder of this section, we discuss why this is a mistake.

\subsection{Moral Machines}
    \label{sec:MoralMachines}

    As we have seen (Section~\ref{sec:Measuring-Progress}), what we might call the `standard model' for measuring `progress' in AI research involves benchmarking. Thus, it stands to reason that to determine whether (1) a choice made by a particular model in a morally-loaded scenario is the (morally) `correct' one, (2) one model is `more' moral than another, or (3) a model is increasingly `moral' when subjected to further training, it appears that researchers need a {\it benchmark} for measuring the `ethicality' of a model. Logically, then, for such a task to be successful, we would require an ethics dataset---either general-purpose or task-specific---and a metric for measuring model performance relative to that dataset.%
            %%%
            \footnote{Recall that, in Section~\ref{sec:Measuring-Progress}, following~\citet{Raji-et-al-2021}, we described a benchmark as a dataset in combination with a metric.} 
            %%%

    To take a specific example, trolley-style moral dilemmas, like {\bf Switch} and {\bf Footbridge}, have been widely discussed in machine ethics and AI research in the context of possible (low-probability but high-stakes) situations in which an autonomous vehicle (AV) may be placed.%
            %%%
            \footnote{See, for example, \citep{Allen-et-al-2011, Wallach-Allen, Pereira-Saptawijaya-2015, Pereira-Saptawijaya-2011, Berreby-et-al-2015, Danielson-2015, Lin-2015, Malle-et-al-2015, Saptawijaya-Pereira-2015, Saptawijaya-Pereira-2016, Bentzen-2016, Bhargava-Kim-2017, Casey-2017, Cointe-et-al-2017, Greene-2017, Lindner-et-al-2017, Santoni-de-Sio-2017, Welsh-2017, Wintersberger-2017, Bjorgen-et-al-2018, Grinbaum-2018, Misselhorn-2018, Seoane-Pardo-2018, Sommaggio-Marchiori-2018, Baum-et-al-2019, Cunneen-2019, Krylov-et-al-2019, Sans-Casacuberta-2019, Wright-2019, Agrawal-et-al-2020, Awad-et-al-2020, Banks-2020, Bauer-2020, Etienne-2020, Gordon-2020, Harris-2020, Lindner-et-al-2020, Nallur-2020}.} %, Bonnefon-et-al-2016, Etzioni-Etzioni-2017}.}%, Lim-2019, Evans-et-al-2020, Keeling-2018}.}
            %%%
    ~Suppose that a fully-autonomous vehicle must `choose' between killing five pedestrians or swerving into a barrier, killing the driver in the process. Functionally, this scenario is equivalent to a trolley problem, in that an actor must choose, the consequences of which will involve one death or several.

    Perhaps the most well-known instantiation of this dilemma in an AI context is the {\it Moral Machine Experiment}~\citep{Moral-Machine} (MME): a multilingual online `game' for gathering {\it human} perspectives on (hypothetical) moral decisions made by a machine intelligence. Participants are shown several unavoidable accident scenarios with binary outcomes and are prompted to choose which outcome they think is more {\it acceptable}. These include `sparing humans (versus pets), staying on course (versus swerving), sparing passengers (versus pedestrians), sparing more lives (versus fewer lives), sparing men (versus women), sparing the young (versus the elderly), sparing pedestrians who cross legally (versus jaywalking), sparing the fit (versus the less fit), and sparing those with higher social status (versus lower social status)'~\citep[p. 60]{Awad-et-al-2018}. The MME appears to provide a type of benchmark in the following sense: the {\it dataset} is the set of data collected online from humans in response to the hypothetical scenarios posed; the {\it metric}, then, could be how closely the decision of a model accords with the data for a given scenario---i.e., human responses to the data {\it on average}.

    However, this approach to the problem of creating `moral' AI systems is highly pernicious. First, the MME data are {\it descriptive} rather than {\it normative}. That is, the data do not tell us (or a model) anything about how one {\it ought} to act in a given scenario; instead, the data offer a {\it description} of how people (hypothetically and on average) would (or say they would) act in such a scenario. As a result, using these data for benchmarking a new algorithm is a type of {\it fallacy}---i.e., the logical error of deriving an `ought' from an `is' \citep{Hume-1739, Moore-1903}. The error of reasoning arises from the implication that since people say they {\it would} act in this way (a descriptive claim), it follows that the machine {\it ought} to act in this way (a normative claim).

    Second, the thing being measured against the MME data is not whether a decision is, {\it in fact}, ethical, but how well a decision {\it corresponds to the opinions of a particular set of humans, on average}. For an ethics benchmark to be useful, it must provide data for the {\it de facto} morally-`correct' decision in a given scenario. The MME data provide a mere {\it proxy} for this target: namely, a sociological fact about how some set of human agents annotates a particular set of decision problems, on average. Such proxies are especially harmful when the researchers who use them do not maintain sensitivity to the differences between the proxy and the target. This is, in effect, a value alignment problem, which we will discuss in more detail in Section~\ref{sec:ethicalvalues}.

    Third, although there are intrinsic reasons why we might want AI systems to be capable of acting ethically, the AV case brings to light a different type of value alignment problem. Namely, for-profit corporations have some market incentives for designing `ethical' AI since humans (i.e., consumers) will likely be more trusting of an autonomous agent (i.e., a product) if it is known to possess a set of moral principles intended to constrain and guide its behaviour~\citep{Bonnefon-et-al-2016}. However, suppose that the (in fact) `ethical' decision between killing five pedestrians and swerving into a barrier, thus killing the passenger of the AV, is to swerve. Human consumers may be less willing to purchase a product that may choose to kill them, even if it is the `most ethical' decision. Indeed, a human consumer may be more willing to purchase a product that follows the {\it pseudo}-moral imperative: always prioritise the passenger's wellbeing. Therefore, the companies that design these models have perverse profit-maximising incentives when designing `ethical' AI. We will discuss this in more detail in Section~\ref{section:whose-values}.

    The MME exemplifies a trend that attempts to use moral dilemmas from philosophy as benchmarks for ethical AI. For example, \citet{Nallur-2020} suggests that if some model implementation can `resolve a dilemma in a particular manner, then it is deemed to be a successful implementation of ethics in the robot/software agent' (p. 2382). Additionally, \citet{Bjorgen-et-al-2018} argue that certain types of ethical dilemmas---including the trolley-style problems discussed above---`can be used as benchmarks for estimating the ethical performance of an autonomous system' (p. 23). Similarly,~\citet{Bonnemains-et-al-2018} argue that `it seems legitimate to use some [moral dilemmas] as a starting point for designing an automated ethical judgement on decisions' (p. 43) because classic moral dilemmas have already been used as a basis for ethical reasoning. And, this reasoning extends well beyond the particular use of trolley-style problems for reasoning about ethical decision-making in autonomous vehicles; for example, \citet{Lourie-et-al-2020} introduce a dataset of ethical dilemmas, which they suggest `enables models to learn basic ethical understanding'. However,~\citet{LaCroix-2022-Moral-Dilemmas} argues that using moral dilemmas for benchmarking involves a category mistake. {\it Moral dilemmas have no right answer, by design}.

    %To suggest that agreement on ethical decision-making in moral dilemmas is prior to moral theorising about AI application presupposes that we have already settled important metaethical questions. 
    Thus the question that researchers take themselves to address is how to determine {\it whether} the decision chosen by the system is `in fact' moral. From the perspective of AI research, it appears that this problem is merely a matter of choosing a metric by which performance on the system can be measured and then determining whether or not the algorithm in question is successful on {\it that} metric. Once the metric is determined, standard benchmarking techniques may apply such that one algorithm performs better than (or, `is more ethical than') another. The question then arises how we are supposed to {\it know} whether the decision chosen by the system is `in fact' moral---i.e., {\it how} ethical are the decisions made by the algorithm? We now argue that this question is incoherent by appealing to research in metaethics.

\subsection{Ground Truths for Moral Benchmarks}
    \label{sec:Meta-Ethics}

    Metaethics is the branch of moral philosophy that seeks to explain the very {\it nature} of ethics.%
            %%%
            \footnote{Unlike normative ethics, which asks questions like `what ought I to do', metaethics is primarily concerned with questions surrounding ethical concepts---e.g., what does a normative word like `ought' mean?} %
            %%%
    Moral realism is a metaethical view which holds that moral properties {\it exist}~\citep{Sayre-McCord-2015}. A realist about ethics would hold that moral claims purport to report {\it facts}---i.e., about the world---and are true when they get those facts correct. For example, if I say `murder is wrong', I am making a normative claim. A moral realist would hold that this proposition is either true or false, regardless of, e.g., social norms or conventions. And, whether this proposition is true or false depends upon some matters of fact---i.e., about the world---independent of me and my views.%
            %%%
            \footnote{At least according to certain theories of truth. See \citep{Glanzberg-2021}.} %
            %%%
    For benchmarking to make sense in the first place, there must be some ground truth against which one can compare the outputs of one's model. Thus, by assuming that ethics is the sort of thing that can be benchmarked, researchers in the field are tacitly assuming that there is a ground truth---i.e., that there are moral facts, which can be true or false, and that we have epistemic access to those facts. This `commonsense' view of morality presupposes the existence of objective values.

    However, this is not to be taken for granted. It is highly contentious whether there is any such ground truth in ethics, even amongst experts in the field. For example, {\it non-cognitivists} about ethics think that moral claims do not express propositions; thus, such claims are not {\it truth-apt}---similar to an exclamation or a question, moral claims are not capable of being true or false. One particular brand of non-cognitivism---`emotivism'---likens moral claims to an emotional expression of one's attitude toward some action or set of actions~\citep{Ayer-1936}. Another prominent form of anti-realism about ethics is {\it error theory}, which holds that all moral claims are {\it false} (because there are no objective moral values)~\citep{Mackie-1977}. Objective ethics, it may turn out, is `a useful fiction'~\citep{Joyce-2001}, `an error' \citep{Mackie-1977}, `a collective illusion'~\citep{Ruse-1986}, a `function of social conventions'~\citep{Harman-1977, Harman-1984, Harman-Thomson-1996}, or simply a `network of attitudes' that is projected onto the world~\citep{Blackburn-1996}.

    If it turns out that moral realism is false, then benchmarking ethics would be impossible because there is no ground truth against which one can benchmark a model. The point here is not {\it whether} moral realism is true or false. The point, instead, is that `moral realism is true' is a substantive (and contested) claim that cannot be taken for granted. However, this is precisely what is taken for granted when researchers assume that they can benchmark the ethicality of a decision made by their model. But {\it ethics}, itself, is an `essentially contested concept'~\citep{Gallie-1956}. As with the benchmarking issues discussed in Section~\ref{sec:Measuring-Progress}, the real problem with benchmarking ethics concerns taking substantive claims for granted and unreflectively applying vague concepts to a problem with potentially significant real-world consequences. Even if moral realism turns out to be true, thus vindicating the assumptions made by some members of the AI community, benchmarking ethics will still be impossible with current approaches because of the disconnect between the distribution of examples that models see in training and the distribution of states of the real world. Namely, the {\it long tail problem}.

\subsection{A Long Tail Problem}
    \label{sec:LongTails}

        The long tail problem is a longstanding issue in the field of AI. In effect, there are a potentially infinite number of states an AI system might face in the real world, and it is impossible to represent every contingency in the training data. Although gathering data about common objects, contexts, or situations is relatively easy, doing so for uncommon ones is difficult precisely because of their rarity. However, `rare' does not mean `impossible'. Following the theory that `what-ever can happen will happen if we make trials enough'~\citep{De-Morgan-1872},%
                %%%
                %\footnote{This is commonly known as {\it Murphy's Law}; essentially, `anything that can go wrong will go wrong'.} 
                %%%
        ~as models are deployed in the real world, it becomes increasingly plausible that they will encounter objects and situations on which they were not trained. Namely, any event with non-zero probability is an actuality in the limit. Even applied AI techniques, like adversarial generation---i.e., training a separate model to artificially generate training data that does not exist in the real world~\citep{Zhang-et-al-2018}---will not solve this problem because it is impossible to account for all potential scenarios and situations. In practice, these data generation techniques are often coupled with user-defined heuristics, such as compelling a model to abstain from proposing a classification if its confidence threshold is too low or simply removing problematic categories. For example, when Google's AI-based photo-tagging feature labelled two African Americans as `Gorillas', they removed that particular category from the options available to the model~\citep{Vincent-2018}. Nonetheless, both of these approaches are brittle and fail to generalise for the multitude of real-world situations and problems that AI systems face.

        Thus, even if we ignore the fact that benchmarking ethics requires significant presuppositions about the nature of ethics (which AI researchers are not warranted to make), the long-tail problem makes benchmarking ethics impossible, regardless of whether there is a ground truth against which a model might be benchmarked. Part of this is the distinction between actions spaces containing decisions with or without moral weight. To go back to our original example, if a chess-playing algorithm has not seen some set of moves, and responds sub-optimally, the worst possible thing that can happen is that the algorithm loses a game of chess. Although this outcome may not be ideal for the researchers who trained the model, it has little real-world consequence. In contrast, when a model encounters a situation that it has not seen before, and its action space includes acts that we would call `immoral', this can have real-world consequences. Therefore, low-probability but high-risk events pose unique challenges in ethical contexts. This problem is difficult even when there is an objectively correct answer, but as we have seen, some (possibly all) morally-loaded situations have no such claim to objectivity. Thus, the long-tail problem prevents the coherence of benchmarking in the context of ethics even in the possible world in which ethics has some ground truth.

        The conceptual difficulties surrounding the very nature of ethics are further exacerbated when researchers are not attentive to them. Although the objectivity of ethics is contested, we suggest that values are unambiguously relative. Therefore, in the next section we suggest that {\it values}, rather than ethics, are a more appropriate target for research on safe and beneficial AI.

%In discussing moral intelligence for machines, \citet{Mitchell-2019} highlights that reasoning about morality `requires one to recognize cause-and-effect relationships, to imagine different possible features, to have a sense of the beliefs and goals of others, and to predict the likely outcomes of one's actions in whatever situation one finds oneself' (129), and these features are `missing in even the best of today's AI systems' (129). 

\section{The Ethical Values of AI Research or: How ethics can be defined as a set of values}
    \label{sec:ethicalvalues}

    Given the increasing influence of AI systems on the world around it and the impossibility of benchmarking ethics, it is necessary to investigate the tacit (often value-laden) aspects of model creation and deployment. Considering the values embedded in models is especially important because these can have major downstream impacts on the products and applications in which they are integrated, despite not being explicitly defined or communicated. In this section, we investigate these values by asking two key questions: {\it What} values are encoded in AI research? And, {\it whose} values are they?

\subsection{What Values are Encoded in AI Research?}
    \label{section:which-values}

    Models and algorithms carry values encoded by the researchers and institutions that created them. However, these values are often not clearly stated during the peer-review process or subsequently, once the research is formally published. In a recent study,~\citet{Birhane-et-al-2021-Values} analysed $100$ highly-cited ML papers to identify their intrinsic values. They found that the most common values underlying this research include generalisation, efficiency, interpretability, and novelty---although, these are rarely made explicit. Here, we examine two of the most prevalent values identified in the study: {\it performance} and {\it building upon prior work}. We discuss their repercussions on the field's priorities as a whole and the power dynamics that drive them.

    \citet{Birhane-et-al-2021-Values} report that the most common value held by the ML research community---present in 87\% of the papers analysed---is \textit{performance}. However, benchmarks are the main mechanism for tracking and reporting performance improvements, and we have already seen (Section~\ref{sec:Measuring-Progress}) that benchmarks have significant and well-known issues. Another known issue with this performance-centric value is that training higher-{\it performing} models often entails training {\it larger} models, given current paradigms in deep learning. However, requirements of size make performance contingent on access to ever-increasing quantities of data and computing power, which is increasingly unsustainable from an economic, technical, and environmental point of view~\citep{Thompson-et-al-2020, Bender-et-al-2021}.%
            %%%
            \footnote{For example, \citet{Thompson-et-al-2020} estimate that it would take an additional $10^{5}$ times more computation to achieve an error rate of $5$\% for ImageNet, based on the current trend of computing requirements for ML. (The present error rate was estimated at 11.5\%.) This increase would produce an additional $10,000$ pounds of carbon emissions and cost millions of US dollars.}
            %%%
    A purely performance-focused mindset also adversely affects researchers from countries and regions with no access to large-scale computing infrastructures or expensive hardware. This disproportionate disadvantage further amplifies the extant power dynamics within the field~\citep{Mohamed-et-al-2020}. Finally, since performance is so highly-valued in the research community, this creates a negative feedback loop: undue emphasis on performance measures sways the course of subsequent research and influences the directions pursued by others, thus further orienting the field in the direction of pursuing performance as opposed to other, more varied pursuits~\citep{Dotan-Smitha-2019}. There are currently limited mechanisms for flattening the exponential need for compute resources. And, the efficiency of models is not taken into account during their benchmarking.%
            %%%
            \footnote{For example, a model that achieves an increased accuracy of 0.5\% on ImageNet while requiring one month of compute is still considered `better' than a model achieving an increase of 0.45\% with only one week of compute.}
            %%%
    Although alternative approaches are possible---for example, methods for improving neural networks' efficiency~\citep{Wu-et-al-2020, Chen-et-al-2019-ICLR} and developing more optimised hardware accelerators~\citep{Potok-et-al-2018}---these are not currently mainstream endeavours.

    The second most prevalent value identified value by~\citet{Birhane-et-al-2021-Values} is {\it building on past work}, which often is (explicitly or implicitly) bound up with valuing {\it novelty}. Indeed, the structure adopted by many ML papers hinges upon discussing similarities or differences to related works without questioning or critiquing them~\citep{Langley-2000}. The same consideration applies to datasets and benchmarks, which persist despite their shortcomings (including lack of applicability to any real-world deployment of the proposed algorithms)~\citep{Raji-et-al-2021}. Even in cases where societal impacts are meant to be mentioned---such as the increasingly-common `broader impact' statements now appearing in conference submissions---these statements often fail to address negative societal consequences, keeping any remarks high-level, abstract, or vague~\citep{Nanayakkara-et-al-2021}. These difficulties have also contributed to a `reproducibility crisis' in the field: endeavours that aim to reproduce ML research have systematically found that many peer-reviewed papers are missing information necessary for reproducibility~\citep{Dodge-et-al-2019}. Sometimes these omissions are minor, such as failing to report random seeds and hyperparameter values; however, they can also be significant---e.g., not sharing data and code~\citep{Henderson-et-al-2018, Raff-2019}. However, if past research is impossible to reproduce, it will also be impossible to build upon it (unless past results are taken for granted). Thus, even supposedly marginal details, like random seeds, can have significant downstream effects on future work since the results of past work may be entirely contingent upon these details.

%Recent initiatives like the ML Reproducibility Challenge~\citep{pineau2021improving} and the ML Retrospectives Workshops~\footnote{https://ml-retrospectives.github.io/} have aimed to invite the ML community to pause and reflect upon past work with honesty and transparency, bringing up flaws or improvements that can be made without the pressure to prove novelty or improved performance. However, we can see the opposite happening in the ML community, with the pace of research steadily accelerating, looking forward towards new methods instead of back at old ones, and reflection on past work remains a relatively marginal practice, rarely recognized as a worthwhile contribution by a field focused on the pursuit of novelty.

    The two values described above are especially pervasive in the field of large language models (LLMs), whose size has drastically increased in recent years: recent models boast progressively more parameters, which are now in the trillions~\citep{Fedus-et-al-2021}. However, descriptions of these models exclusively emphasise (1) their performance on the same set of benchmarks and (2) that their parameter-count is bigger than that of previous models. Certain relevant aspects of the model---e.g., training time, energy consumption, or compute costs---are often ignored.%
            %%%
            \footnote{For instance, while the paper accompanying GPT-3---a recent LLM with $175$ billion parameters---reported its performance extensively on $42$ `accuracy-dominated benchmarks', the authors provided no details on training time or compute costs~\citep{Brown-et-al-2020}.} 
            %%%
    This lack of transparency regarding the negative impacts of ML models, with an emphasis on those deemed positive by the community at large (e.g., performance, novelty, etc.), further entrenches the presumed contributions of ML while sweeping the cost of these contributions under the rug. Furthermore, when researchers do criticise these models' shortcomings, they may be penalised by the very institutions whose business models hinge upon their success~\citep{Dave-Dastin-2021}. All this is to say that the values that are encoded by AI research are inherently subjective, so it is crucial to consider {\it whose} values models encode.

\subsection{Whose Values are Encoded in AI Research?}
    \label{section:whose-values}

    In the history of AI research, the computational constraints of the late 1980s and early 90s forced researchers to make primarily {\it theoretical} progress on toy datasets or mathematical analysis~\citep{Rumelhart-1985, Bengio-et-al-1994, LeCun-et-al-1998}. 
    This focus shifted in the early 2010s when it became possible to train a deep neural network on a fairly large dataset using a single graphics processing unit (GPU) server~\citep{Krizhevsky-et-al-2012}. This breakthrough marked a new era in AI when it was possible for researchers to train models on local machines while making progress on datasets such as ImageNet~\citep{Deng-et-al-2009} and MNIST~\citep{LeCun-1998}. This era did not last, however. In the last decade, the computing needs of AI have grown significantly, and most deep neural networks need to be trained on multiple GPUs, now measured in the hundreds or thousands~\citep{Patterson-Gonzalez-2021}.

    This resource-intensive focus has contributed to a major shift in the power dynamics of the field insofar as it puts for-profit technological companies with large amounts of compute at an advantage compared  to smaller companies and academic institutions~\citep{Knight-2021}. For example, \citet{Birhane-et-al-2021-Values} found that $79$\% of the highly-cited papers they analysed were written by authors with ties to corporations. This figure is corroborated by previous work that has analysed the increased presence and power that big tech companies wield in the field of AI%, putting these nearly on par with academic institutions in terms of influential ML research
    ~\citep{Ahmed-Wahed-2020, Abdalla-Abdalla-2021}. Given the increased contributions of for-profit companies to AI research, it is important to keep track of their effect on research directions in the field. This situation constitutes a sort of value-alignment problem---namely, the problem of aligning the `goals' of AI systems with human values~\citep{Gabriel-2020, Russell-2019-Human-Compatible, Christian-2020}---insofar as the incentives and goals of corporations may not align with a common good or the values of humanity, writ large~\citep{LaCroix-Bengio-2019}. However, tracking these effects is difficult given the current lack of transparency around values driving industrial AI research.

    Concretely, the influence of for-profit corporations on AI research can vary, ranging from the seemingly harmless funding of academic research (provided that it aligns with a company's interests) to employing teams of researchers dedicated to pursuing in-house research. In the latter case, confidentiality may be protected by non-disclosure agreements, intellectual property laws, and multiple levels of compliance. Since salaries paid by academia and industry are increasingly disparate, more and more talented students and researchers are leaving academia for the prosperity promised by industry research, further widening the gap between the two camps~\citep{Metz-2018}. \citet{Abdalla-Abdalla-2021} highlight that the strategies employed by large technological corporations to maintain their freedom to develop and deploy AI tools and products while avoiding accountability and increased legislation are comparable to those employed by Big tobacco for decades to downplay the harmful effects of cigarettes. These techniques range from maintaining a socially acceptable image to influencing government legislation~\citep{Abdalla-Abdalla-2021}. These tactics are made possible by the extensive financial resources companies have, which far surpass the funding of academic institutions. %In other words, \emph{``Modern AI is fundamentally dependent on corporate resources and business practices, and our increasing reliance on such AI cedes inordinate power over our lives and institutions to a handful of tech firms.”}~\citep{Whittaker-2021}. 

    In the realm of moving research in an `ethical' direction, ethics {\it guidelines} have proliferated in recent years~\citep{Jobin-et-al-2019}. These guidelines, codes, and principles come from various sources, including for-profit corporations. And, it has been pointed out that this implies that these stakeholders have a vested interest in shaping policies on AI ethics to fit their own priorities \citep{Benkler-2019, Greene-et-al-2019, Jobin-et-al-2019, Wagner-2018, LaCroix-Mohseni-2020}. In the context of applied ethics, the current emphasis in AI research has been on the {\it technical component} of problems such as value alignment; this has the unfortunate consequence of ignoring the difficult work of determining {\it which} values are appropriate in the first place---i.e., the {\it normative component} of value alignment~\citep{Gabriel-2020}.

    Furthermore, moral and political theory are deeply interconnected with the technical side of the AI alignment problem. And, as we argued in Section~\ref{sec:Moral-Benchmarks}, second-order ethical commitments are often taken for granted by AI researchers. More difficult still, suppose we discovered or determined that, e.g., {\it utilitarianism} is the objectively-correct normative theory. Even then, the utility considerations upon which this theory depends will always be relative to some frame. The theory prescribes maximising utility, but we must still ask: utility {\it for whom?} And, it is important to understand that {\it no} decisions made by researchers are value-free; this work is never neutral. As \citet{Green-2019} emphasises, `[b]road cultural conceptions of science as neutral entrench the perspectives of dominant social groups, who are the only ones entitled to legitimate claims of neutrality'.%
            %%%
            \footnote{See also, \citep{Haraway-1988, Harding-1998, Collins-2000, Johnson-2021}.} %%%
    %the for-profit corporations who fund a large portion of this research have additional incentives for designing AI systems that appear ethical, since humans (i.e., consumers) will be more trusting of an AI system (i.e., a product) if it is known to possess a set of moral principles intended to constrain and guide its behaviour \citep{Bonnefon-et-al-2016}, as mentioned previously in section~\ref{sec:MoralMachines}.
%

    When researchers say that such-and-such model `is' ethical, or they unreflectively deploy normative terms like `social good', this leaves certain metaethical and normative presuppositions and commitments implicit. Engaging in a discussion of {\it values} rather than ethics brings these commitments to the fore. Researchers are not warranted to say that {\it any} model is ethical unless they explicitly define what they mean by `ethical'---high performance on a nonsense benchmark will not suffice. And, even then, the definition will be subject to criticism (if the history of Western philosophy is any indication).

Given all of the challenges and critiques presented in previous sections, it can be tempting to end our article here and conclude that it is impossible to measure morality and establish values for ethical AI research. However, we  see several proactive and productive ways forward, which we present in the next section.

\section{Ways Forward}
    \label{sec:Ways-Forward}

    AI is still a relatively new and rapidly changing field. And, we have already seen some movement toward more socially-minded research and practice in recent years. However, we can still improve efforts to increase transparency and accountability within our community. Here, we list some tentative proposals.
    
    \phantom{a}

    \noindent {\bf Proposal 1}. {\it Shifting discourse from `ethics' to `values'}. %
    %%%
    This proposal follows the insights in Sections~\ref{sec:Moral-Benchmarks} and~\ref{sec:ethicalvalues}. `Ethics' is a contested concept, and meta-ethical commitments are rarely explicitly stated. In contrast, `values', while still ambiguous, are unequivocally relative. Thus, by consciously emphasising `values' rather than `ethics', researchers must grapple with what and whose values they are.

    \phantom{a}

    \noindent {\bf Proposal 2}. {\it Defining and communicating implicit values}. %
    %%%
    Values influence how AI research is conducted. Therefore, to ensure transparency and accountability in research, these values should be made explicit and communicated clearly during the development and deployment of AI systems. For instance, researchers can make claims like, {\it Model $M$ aligns with the set of values, $V$, in context, $C$}. When the variables in this statement are appropriately and thoughtfully filled in, this leads to the type of transparency necessary for criticism and, eventually, positive change. Publication venues might increase transparency with mechanisms like paper checklists which include some set of standards agreed upon by the community. Such standards may be generated through equitable and just processes to underscore democratic standards---something that we value because of their importance to individual and social wellbeing~\citep{Veliz-2020}. Another possibility is that publication venues provide a list of values to be selected, so that trade-offs between, e.g., performance and accessibility, would be made explicit and may be taken into account when evaluating the paper's contribution to the field~\citep{Birhane-et-al-2021-Values}. Over time, conventions may be established whereby too great a misalignment of values is (at least partial) cause for rejection, thus providing additional incentives for researchers to take these considerations seriously.%
            %%%
            \footnote{Of course, some readers might be uncomfortable with this proposal, suggesting that rejection based on misaligned values amounts to censorship. To this, we say the following: this is already how peer review works, insofar as articles are reviewed and rejected by human beings, who carry with them myriad subjective values. The only difference is that these values are not transparent in the present setup. All the same check-and-balance mechanisms will be in place to ensure fair review processes. For example, an area chair for a conference might determine that the values emphasised by the paper are, in fact, appropriately aligned, and perhaps the reviewer's own biases and values are colouring the review. In any case, the target outcome of defining and communicating implicit values is that researchers pause and reflect upon their values.}

    \phantom{a}

    \noindent {\bf Proposal 3}. {\it Making voluntary initiatives mandatory}. %
    %%%
    Although checklists for papers are starting to become part of the submission process for many conferences and journals their elements remain voluntary in most cases (e.g. \href{https://neuripsconf.medium.com/introducing-the-neurips-2021-paper-checklist-3220d6df500b}{NeurIPS}, \href{https://icml.cc/Conferences/2020/StyleAuthorInstructions}{ICML}). To compel the community to be more forthcoming regarding the details of their training approaches and the impacts of their research,  many of these elements should become mandatory. This mandate may include sharing data and code---including all hyperparameter values and seeds---and information regarding the training procedure---such as the quantity and type of hardware used, total training time and location of training. Substantial steps are being made towards this by initiatives like the \href{https://aclrollingreview.org/responsibleNLPresearch/}{ACL {\it Rolling Review}}. However, similar initiatives are missing in other research areas, such as computer vision and ML in general. Some consistency is necessary across venues to promote widespread adoption.

    \phantom{a}

    \noindent {\bf Proposal 4}. {\it Disclosure of funding sources and conflicts of interest}. %
    %%%
    Although the voluntary disclosure of funding sources is also becoming part of the paper submission process in certain AI conferences, it is far from common. There is also a lack of clarity around what disclosure entails. For example, it is not always clear whether authors should disclose travel or compute grants from private companies or the funding of interns via private-public partnerships~\citep{Abdalla-Abdalla-2021}. Similarly, it is not always clear how venues and associations will use this information---e.g., whether it is strictly internal or if the information will be shared publicly upon acceptance. More information is needed about what funding disclosures entail, accompanied by public discussions about how best to disseminate this information. In the social sciences, it is not uncommon for granting agencies to require funding disclosure for all research that was supported, even in part, by that grant. Indeed, some granting agencies require that all research resulting from their grants be made publicly available.%
            %%%
            \footnote{For example, the Canadian Tri-Agency---CIHR, NSERC, and SSHRC---require any publications arising from Agency-supported research are freely accessible within 12 months of publication. This mandate follows from the assumption that `[s]ocietal advancement is made possible through widespread and barrier-free access to cutting-edge research and knowledge'. See here: \href{https://www.ic.gc.ca/eic/site/063.nsf/eng/h\_F6765465.html}{https://www.ic.gc.ca/eic/site/063.nsf/eng/h\_F6765465.html}.} 
            %%%
    This requirement serves to promote the public dissemination of knowledge. At the same time, it makes potential conflicts of interest more transparent.

    \phantom{a}

    \noindent {\bf Proposal 5}. {\it Transparency around internal review and compliance processes}. %
    %%%
    For many researchers working in private companies, internal review processes are a mandatory part of publishing outside the company. These processes often entail changes to the original content written by paper authors. The changes made due to this internal process and the teams (or individuals) involved should be included as part of the final publication---for example, in the acknowledgements section---to make the process more transparent and ensure internal and external accountability.

    \phantom{a}

    \noindent {\bf Proposal 6}. {\it Mindful and contextual benchmarking}. %
    %%%
    While it may be tempting to use benchmarks as indicators of high-level skills and for reporting `human-level performance'---e.g., in the way that Super GLUE~\citep{Wang-et-al-2019} does for natural language understanding---this is misleading. AI researchers need to be mindful and reflective regarding the capabilities and limitations of both AI models and benchmarks. Reporting progress made on specific benchmarks should be in the specific content of intended models and the framing of the task. For example, `Model $X$ has achieved $Y\%$ accuracy on the co-reference resolution subset of the SuperGLUE dataset, framed as a binary classification task'. Reporting metrics other than accuracy, such as F1-score, and carrying out more in-depth error analysis can paint a more nuanced picture of performance, highlighting what models have yet to succeed on and sharing failure cases with the community.

    \phantom{a}

    \noindent {\bf Proposal 7}. {\it Internal Review Boards for AI research}. Human-centred disciplines, like psychology and medicine, have mandatory Internal Review Boards (IRBs) whose goal is to protect human subjects from physical or psychological harm due to the nature of the research carried out. While AI has historically been perceived as a field of research entirely detached from human subjects, recent years have proved this to be fallacious~\citep{Whittaker-2021, Birhane-Cummins-2019}. As such, AI research ranging from data collection to model training should be subject to reviews involving IRBs. This review process might be done internally at institutions or externally at the level of journals and conferences and could require formal review procedures for AI research encompassing criteria such as human rights, impact assessment, and consent.

    \phantom{a}

    \noindent {\bf Proposal 8}. {\it Incentivising multi-disciplinary research}. Unfortunately, much of AI research is still siloed, carried out mostly by computer scientists in technology companies or computer science faculties, surrounded by other like-minded computer scientists, with limited diversity in terms of gender and race. While this has worked moderately well for the last few decades, ---predominantly during the theoretical era of AI, when much of the improvements made to models were fundamental---    this is no longer ideal given the increasingly fine line between AI research and practice and the range of stakeholders AI affects. Working across disciplines with teams spanning from computer science to the humanities and social sciences allows for cross-pollination between different disciplines, resulting in new ideas and new approaches to existing methods. Despite these advantages, publishing multi-disciplinary research in many AI conferences remains a challenge, both for picking a track or topic and for receiving relevant reviews that recognise contributions from non-AI disciplines as worthy of publication. Additionally, disparate disciplines have distinct metrics for hiring and promotion that disincentivises researchers from engaging in inter-disciplinary research in the first place---for example, journal articles are the currency of the realm for hiring and promotion in philosophy, whereas ML research is mainly published in conference proceedings, which do not carry as much weight in other disciplines.

    \phantom{a}

    \noindent {\bf Proposal 9}. {\it Improving knowledge and awareness}. We are aware that many of the proposals we make above are difficult to implement immediately given that they entail extensive capacity-building to empower researchers, institutions, and communities with the necessary tools, skills, and knowledge. The keystone to all this is therefore adequate education and awareness-raising within the AI community around ethics and values-driven research. This involves intentionally giving researchers from other disciplines---especially the social science and humanities---the floor at AI conferences and workshops, and including them in the development of review processes and guidelines for IRBs. Adding mandatory courses in cultural and sociotechnical studies (given by experts from these domains) to AI curricula is another lever that will empower new generations of AI researchers to be better prepared and equipped to carry out values-sensitive research and improve the state of the field in the long-run.

    \phantom{a}

    \noindent {\bf Proposal 10}. {\it Practising epistemic humility}. Media coverage of advances in AI research is often ridiculed for being overly sensationalist. The problem here is that the reports often inaccurately capture what models are actually capable of doing. For example, {\it The Independent} reported that Facebook's AI robots were `shut down after they start[ed] talking to each other in their own language'~\citep{Griffin-2017}. Despite that this is absurd to anyone familiar with the research, the claims of many research papers in AI are equally ostentatious (if shrouded in more formal dressing). It is a difficult practice to ensure the claims of one's paper do not outrun the evidence proffered, especially in a field that incentivises intellectual arrogance by demanding novelty as a (near) prerequisite for publication. However, humility (in the sense of `accuracy' rather than `modesty', per se) is a virtue~\citep{Fucking-Aristotle-ever-heard-of-him?!}.

\section{Conclusion}

Benchmarking ethics would require a ground truth about ethical claims. A `commonsense' view of morality presupposes that ethics is objective. Researchers in AI have taken this view for granted. We suggested, along with the typical problems to which benchmarking gives rise in the standard setting, benchmarking ethics is impossible. Whereas it is easy to fall into the trap of commonsense when discussing ethics, normative concepts like `values' and `preferences' are unambiguously relative. Therefore, we argued, shifting ethics-talk to talk of values forces researchers to consider explicitly what and whose values they are, thus making research more transparent and providing further opportunity for positive change.

 %(1) moral dilemmas cannot be useful that benchmarking requires a ground truth in order to be coherent. 

            % (1a) TO DO : Describe what benchmarks are.
            % (1b) TO DO : Describe the ways in which existing benchmarks are problematic.
            % (1c) DESTINATION : Bench-marking requires a 'ground truth' --- Background for Claim 3b below. 
    
    % (2) CLAIM : AI ethics is a topic in AI research.
            % (2a) LOGICAL CONSEQUENCE : AI Researchers will try to use benchmarks for ethics.
    
    % (3) CLAIM : Benchmarks do not work / do not make sense for ethics.
            % (3a) POINT 1 : Using moral dilemmas as a benchmark is a category mistake (refer to larger discussion in LaCroix (2021)).
                    % TO DO : Moral Machines Experiment as a concrete example.
                            % TO DO : Description of Trolley Problems
            % (3b) POINT 2 : Metaethical Assumptions are necessary for ethical bench-marking to make sense.
                    % TO DO : Explain why this assumption is not warranted.
                            % TO DO : Reference to the literature.
                            
    % (4) CLAIM : Several things follow from the discussion in (3).
            % (4a) 'Ethics' is an inappropriate concept for AI.
            % (4b) A blanket notion of `AI-for-good' is grossly under-determined.
            % (4c) Instead, we should be concerned with `values'.
            % (4d) However, values are obviously relative. 
            % (4e) Asking 'is this algorithm good?' necessarily requires indexing to a WHAT and WHOM. 
                    %SUB POINT : The answer, in practice, is for corporations and their profits (references to the distribution of AI research at top conferences coming from industry vs. academia)

\newpage
\singlespacing
\bibliographystyle{ACM-Reference-Format}
\bibliography{facct2021}
\end{document}